%% file: ndlib.tex
\documentclass[submission,creativecommons]{eptcs}
\providecommand{\event}{DSL 2011}

\usepackage{graphicx}
\usepackage{fancyvrb}
\usepackage[format=hang]{caption}
\usepackage[hypcap=true]{subcaption}
\usepackage{suffix}

\newcommand{\ttsize}{\small}

\DefineVerbatimEnvironment{code}{Verbatim}%
   {fontsize=\ttsize,commandchars=\\\{\},samepage=true} 
\DefineVerbatimEnvironment{badcode}{Verbatim}%
   {commandchars=\\\{\},samepage=true,frame=single,
    xleftmargin=9em,xrightmargin=9em}
\DefineVerbatimEnvironment{figcode}{BVerbatim}%
   {fontsize=\ttsize,boxwidth=auto,commandchars=\\\{\}}

\newcommand{\cod}[1]{{\ttsize\texttt{#1}}}

\newcommand{\figscale}{0.62}

\input{ndd.tex}

\newcommand{\mysection}[1]{\vspace{-.5ex}\section{#1}\vspace{-0.3ex}}
\newcommand{\mysubsection}[1]{\vspace{-.5ex}\subsection{#1}\vspace{-0.3ex}}

\newlength{\dummylen}

\newcommand{\BSLASH}[1][]{\char`\\#1}
\newcommand{\OBRACE}{\char`\{}
\newcommand{\CBRACE}{\char`\}}
\newcommand{\USCORE}{\char`\_}

\newcommand{\Gen}{\name{Gen}}
\newcommand{\Maj}{\name{Maj}}
\newcommand{\MajE}{\name{MajE}}
\newcommand{\Pvt}{\name{Pvt}}
\newcommand{\Wake}{\name{Wake}}

\newcommand{\Shot}{\name{Shot}}
\newcommand{\Washed}{\name{Washed}}
\newcommand{\Remove}{\name{Remove}}
\newcommand{\Saved}{\name{Saved}}

\newcommand{\major}{\cod{majorOrders}}
\newcommand{\both}{\cod{bothOrder}}
\newcommand{\trump}{\cod{trump}}
\newcommand{\Act}{\cod{Action}}

\newcommand\sref[2]{\hyperref[#1:#2]{\ref*{#1}(\subref*{#1:#2})}}
\WithSuffix\newcommand\sref*[2]{\hyperref[#1:#2]{(\subref*{#1:#2})}}

\sloppypar

\begin{document}

\title{A DSEL for Studying and Explaining Causation}
\def\titlerunning{A DSEL for Studying and Explaining Causation}

\author{
Eric Walkingshaw
\institute{School of EECS \\ Oregon State University}
\email{walkiner@eecs.oregonstate.edu}
\and
Martin Erwig
\institute{School of EECS \\ Oregon State University}
\email{erwig@eecs.oregonstate.edu}
}
\def\authorrunning{E. Walkingshaw and M. Erwig}

\maketitle

\begin{abstract} 

We present a domain-specific embedded language (DSEL) in Haskell that supports
the philosophical study and practical explanation of causation.  The language
provides constructs for modeling situations comprised of events and functions
for reliably determining the complex causal relationships that emerge between
these events.  It enables the creation of visual explanations of these causal
relationships and a means to systematically generate alternative, related
scenarios, along with corresponding outcomes and causes.
The DSEL is based on neuron diagrams, a visual notation that is well
established in practice and has been successfully employed for causation
explanation and research.  In addition to its immediate applicability by users
of neuron diagrams, the DSEL is extensible, allowing causation experts to
extend the notation to introduce special-purpose causation constructs. The DSEL
also extends the notation of neuron diagrams to operate over non-boolean
values, improving its expressiveness and offering new possibilities for
causation research and its applications.

\end{abstract}

\mysection{Introduction}
\label{sec:intro}

Cause and effect are fundamental concepts on which science and society are
built.  But what does it really mean for one event to have caused another, and
how can we determine when \emph{causation} has happened?  Philosophers have
been studying these questions for over 2000 years and it remains an active area
of research even today.
In this paper we present a domain-specific language embedded in Haskell (DSEL)
for working with causation problems and to support causation research.  This
language allows users to model a story or situation and then analyze that model
to determine the causes of events, generate alternative scenarios, and
produce visual explanations of causal relationships.

Causation researchers have developed several notations 
for discussing and reasoning about causation.  The most widely used of these
are \emph{neuron diagrams} \cite{Lewis87}, a domain-specific, visual language
for describing causal relationships between events.  Our DSEL is based on an
extended version of this visual language and is in fact primarily a
sophisticated metalanguage for creating, analyzing, and visualizing programs in
the object language of neuron diagrams.  

\begin{figure}[t]
\centering
\subcaptionbox{\label{fig:orders:major}}
  {\includegraphics[scale=\figscale]{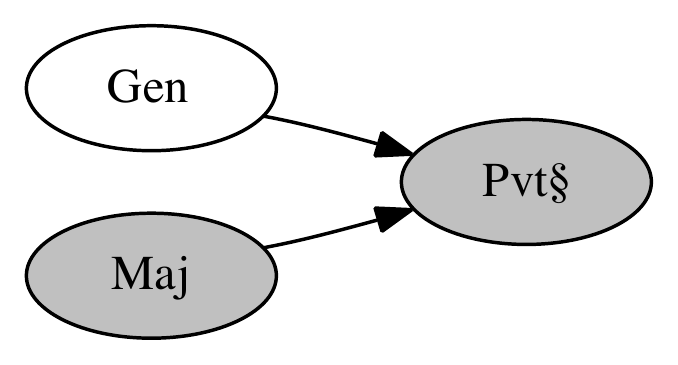}}
\qquad
\subcaptionbox{\label{fig:orders:both}}
  {\includegraphics[scale=\figscale]{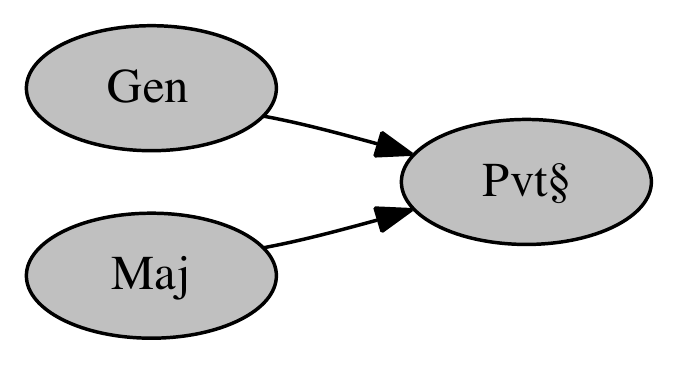}}
\caption[]{Neuron diagrams for a private charging after receiving orders from
superior officers.  In \sref*{fig:orders}{major} only the major has issued an order to
charge, while in \sref*{fig:orders}{both} both officers issue the order to charge.}
\label{fig:orders}
\end{figure}

%

Examples of very simple neuron diagrams can be seen in Figure~\ref{fig:orders}.
Each neuron diagram tells a story.  The stories told by these diagrams (adapted
from an example in \cite{HP05}) concern an army private with two superiors, a
general and a major, each of whom may shout orders which the private will then
carry out.
In the case of the diagram in Figure~\sref{fig:orders}{major}, the general is silent and
the major yells ``charge!'', so the private charges forward.  In
Figure~\sref{fig:orders}{both}, both officers issue the order to charge, and again the
private charges.

A neuron diagram is a directed, acyclic graph (DAG) where each node is called a
\emph{neuron}.  Neurons usually correspond to events in the story and can
either fire or not.  A \emph{firing} neuron is colored gray and indicates the
occurrence of the corresponding event, a \emph{non-firing} neuron is white and
indicates that its event did not occur.
In our examples, the \Gen\ and \Maj\ neurons represent orders to charge by the
corresponding officer.  If the neuron is gray, that officer issues the order to
charge, if the neuron is white that officer issues no order.  The \Pvt%
\footnote{The \lawSym\ symbol adorning the name of this neuron and others
throughout the paper will be explained in Section~\ref{sec:cause}.  It can be
safely ignored for now.}
neuron in each diagram represents the private charging (or not, had the neuron
not fired).

As source nodes in the graphs, the \Gen\ and \Maj\ neurons are called
\emph{exogenous} and their firing values are simply set according to the story.
A downstream neuron, like \Pvt, is called \emph{endogenous}, and its value is
determined by a function on the values of its immediate predecessors.
The function that an endogenous neuron implements is indicated visually by the
shape and style of the node and the style of its incoming edges.  In each of
our examples, the endogenous \Pvt\ neuron is a \emph{standard} neuron
(indicated by its oval shape and standard line thickness) connected to its
predecessors by \emph{stimulating} edges (indicated by triangular arrowheads).
If a standard neuron is stimulated by at least one firing neuron, it will fire.
In other words, each \Pvt\ neuron implements a logical disjunction of its
inputs.

The following code in our DSEL defines the neuron diagram in
Figure~\sref{fig:orders}{major} and binds it to the Haskell identifier \major.

\begin{code}
\major = diagram [pvt] `WithInputs` [False,True]
  where
    gen = "Gen" :# Input
    maj = "Maj" :# Input
    pvt = "Pvt" :# StimBy [gen,maj]
\end{code}

\noindent
%
%
This code will be explained in depth in the next section.  Here we instead
offer a preview of a few things we can do with our language once we have
described a neuron diagram in this way.  First, we can determine the firing
state of any neuron in the diagram by evaluating an expression like the
following (for example, in GHCi).

\begin{code}
> "Pvt" `stateIn` \major
True
\end{code}

\noindent
The value \cod{True} corresponds to a neuron firing, and we can confirm this
result by noting that the \Pvt\ neuron is colored gray in
Figure~\sref{fig:orders}{major}.

Second, we can generate the visual representation of the neuron diagram shown
in the figure.  One way to do this is to evaluate the expression \cod{view
\major}, which generates a GraphViz\footnote{\url{http://www.graphviz.org}}
program that draws the image and loads it in an appropriate viewing
application.  From here the image can be exported and posted on the web,
attached to an email, or included in a paper.  The images in
Figure~\ref{fig:orders}, and in fact all of the images used in this paper, were
generated by our language in this way.

Third, we can use the definition of \major\ to derive diagrams for alternative,
related scenarios.  For example, to produce the diagram in
Figure~\sref{fig:orders}{both}, in which both officers issue the order to
charge, we simply change the input values to our existing diagram as shown
below.

\begin{code}
\both = \major `changeInputs` [True,True]
\end{code}

\noindent
We bind the result to the identifier \both\ so that we can refer to it later.

Finally and most importantly, we can analyze the causes of the terminal neurons
in the diagram.  
In the simple scenario depicted in Figure~\sref{fig:orders}{major}, it seems
obvious that the major's order caused the private to charge.  To confirm, we
can evaluate the following expression and examine the result.

\begin{code}
> causes \major
Maj:True ==> Pvt:True
\end{code}

\noindent
This output indicates that because the \Maj\ neuron fired, the \Pvt\ neuron
also fired, confirming our intuition.  We can compare this to the causes
identified in the diagram bound to \both, shown below.

\begin{code}
> causes \both
Gen:True | Maj:True ==> Pvt:True
\end{code}

\noindent
This result indicates that either the general's order or the major's order is a
sufficient cause of the private's actions.  At first this seems to agree nicely
with intuition---it doesn't matter who issues the order to charge, as long as
one of them does---but as we'll see in Section~\ref{sec:model}, this outcome is
highly dependent on our modeling of the scenario.  Many alternative models with
the \emph{same outcomes} produce \emph{different causes} that could also be
considered correct.

The causal analyses performed above are the product of our own previous
theoretical work on neuron diagrams \cite{EW10hcc}.  Although neuron diagrams
have become very popular in the philosophical community, they have become so in
spite of several (perceived and actual) technical limitations.
Their main advantages over competing languages are largely related to
usability: they are simple, direct, and highly extensible.  Causal
relationships between events are shown explicitly as edges between nodes,
whereas this information is represented only indirectly in textual languages
(including our own DSEL).
Also, new types of neurons that implement different functions can be invented
on demand for use in a particular story.  While this makes the notation concise
and very flexible, critics have argued that it also makes neuron diagrams too
ad hoc and difficult to reason about~\cite{Hitchcock07}.
Our previous work addresses these concerns with a formal description of neuron
diagrams that transparently accommodates this extensibility.  In this paper we
discuss the implementation of this model and its impact on the design and use
of our DSEL.

Neuron diagrams differ from other causal modeling tools (such as those used in
AI \cite{Pearl09}) in that their primary purpose is not to \emph{solve}
causation problems, but rather to share and \emph{explain} causal situations.
As such, the language has historically been somewhat imprecise.  In addition to
formalizing the language, our previous work also introduces a small extension
to the language that allows neuron diagrams to more precisely model certain
causal situations, and the first cause inference algorithm for neuron diagrams.
Both of these technical improvements are summarized in Section~\ref{sec:cause}
and incorporated in the DSEL.
By providing an implementation of these features, we make it possible, for the
first time, to automatically confirm that a neuron diagram encodes its intended
causal relationships.
%

Finally, this paper introduces a new theoretical contribution to neuron
diagrams, extending the language to operate on non-boolean values.  That is,
neurons in our DSEL may not only fire or not fire, but may take on any
value in an arbitrary finite set.
%
%
We present examples using this extension in Section~\ref{sec:nonbool}.

We expect users to be able to use the DSEL for simple applications with only
minimal knowledge of Haskell.  Such applications include creating neuron
diagrams with standard components and analyzing or drawing existing neuron
diagrams; we present several examples of such applications in the next two
sections.
More advanced applications of the language, such as defining new types of
neurons or extending the language to new types of values, will require
increasingly deeper knowledge of both the DSEL and the host language.  We
describe aspects of the language relevant to these uses in
Section~\ref{sec:rep} and provide examples of such applications in
Sections~\ref{sec:nonbool} and \ref{sec:expl}.
This is followed by a discussion of related work in Section~\ref{sec:rw} and
conclusions and future work in Section~\ref{sec:fw}.

\mysection{Neurons, Diagrams, and Graphs}
\label{sec:basics}

In this section we introduce the basic causal modeling constructs of our DSEL
and further our introduction of the notation and associated terminology of
neuron diagrams.

We begin by more closely examining the DSEL code from the previous section, in
particular the definition of \major, used to generate the neuron diagram in
Figure~\sref{fig:orders}{major}.
Note that we define each of the diagram's three neurons individually in the
definition's \cod{where}-clause---this is not strictly necessary, of course,
but leads to definitions that are easy to read and extend, and so we consider
it the preferred concrete syntax for our DSEL.

Each neuron is a value of type \cod{N~a} (defined below), where \cod{a}
represents the type of values the neuron can take on.  For example, the neurons
in \major\ can either fire or not fire, and so have type \cod{N~Bool}.
The DSEL provides several basic operations for querying neurons, implemented as
Haskell functions.  Some of these are summarized in Figure~\ref{fig:nquery}.

\begin{figure}[t]
\centering
\begin{tabular}{ll}
\hline
\cod{name~~~~~::~N~a~->~Name}  & the name of the neuron \\
\cod{isInput~~::~N~a~->~Bool}  & is the neuron an input neuron? \\
\cod{isExo~~~~::~N~a~->~Bool}  & is the neuron exogenous? \\
\cod{isEndo~~~::~N~a~->~Bool}  & is the neuron endogenous? \\
\cod{preds~~~~::~N~a~->~[N~a]} & the immediate predecessors of the neuron \\
\cod{upstream~::~N~a~->~[N~a]} & all recursive predecessors of the neuron \\
\hline
\end{tabular}
\caption{Basic querying operations on neurons.}
\label{fig:nquery}
\end{figure}

Values of type \cod{N~a} are constructed with the neuron constructor \cod{:\#}.
The left argument to this constructor is a uniquely identifying name (within
the diagram) for the neuron, while the right argument is a \emph{neuron
description}.  A neuron description captures several important properties of a
neuron, including the function the neuron implements, the visual style of the
neuron, and the incoming edges from the neuron's immediate predecessors.

The creation of neuron descriptions constitutes a sort of mini-DSL within the
larger DSL for neuron diagrams.  Values in this language are captured by a type
class \cod{ND}.  A single neuron diagram can contain neuron descriptions of
many different types, as long as they all instantiate this type class---this is
crucial to support the kind of ad hoc extension to the visual language of
neuron diagrams that is common in the philosophy research.
To support this in our DSEL, we locally quantify \cite{LO94} the neuron
description type parameter \cod{d} in the following definition of the neuron
type, \cod{N~a}.

\begin{code}
type Name = String
data N a = forall d. ND d a => Name :# d a
\end{code}

\noindent
We can observe in this definition that \cod{ND} is a multi-parameter type
class, that the first argument \cod{d} is a type constructor, and that the
second argument \cod{a} is the value type of the neuron.
However, we postpone the full definition and explanation of neuron descriptions
until Section~\ref{sec:rep}, after we have presented the necessary background
information.
Here we focus instead on the concrete syntax of neuron descriptions.
In the example from the previous section, we describe two different kinds of
neurons.
First, we define two exogenous input neurons, \Gen\ and \Maj, using the
\cod{Input} neuron description.  Then we use the \cod{StimBy} construct to
describe the \Pvt\ neuron as a standard endogenous neuron stimulated by both of
the input neurons.
We will see many other types of neuron descriptions in the following examples.

The \cod{diagram} \ldots\cod{`WithInputs`} \ldots\ construct, used in the example, creates a neuron
diagram given a list of terminal neurons and a list of values to assign to the
input neurons.  Values are assigned to input neurons in the order
that they are encountered in an in-order traversal of the diagram, starting
from the terminal neurons.  In the definition of \major, \cod{False} will be assigned to the
\Gen\ neuron, and \cod{True} to the \Maj\ neuron.

The assignment of values to inputs is separated from neuron descriptions
because we often want to reuse the causal structure of a diagram while
assigning different values to the input neurons.
Recall that this is how we generated the related diagram in which both officers
gave orders, \both, shown in Figure~\sref{fig:orders}{both}.
We call the underlying causal structure of a neuron diagram a \emph{neuron
graph} \cite{EW10hcc}, and it is captured in our DSEL by the following
straightforward type.

\begin{code}
newtype G a = Graph [N a]
\end{code}

\noindent
As with the \cod{diagram} construct, the list of neurons wrapped in this
type are the terminal neurons of the graph---other neurons in
the graph can be accessed, for example, with the \cod{upstream} function.
In fact, the \cod{diagram} keyword is just a synonym for \cod{Graph},
intended to make diagram definitions read more naturally.

A neuron diagram is then just a neuron graph combined with a list of values to
assign to the input neurons.  This is captured in the DSEL by the following
type.

\begin{code}
data D a = WithInputs (G a) [a]
\end{code}

\noindent
The constructor of this data type is designed to be readable when used as an
infix operation, and forms the second part of the
\cod{diagram} \ldots\cod{`WithInputs`} \ldots\ construct.

%

We can access the underlying graph of a diagram with the function \cod{graph}.
Using this, we can also implement the \cod{changeInputs} operation, that replaces
the input values of a neuron diagram (used to derive the \both\ diagram from
the \major\ diagram in the previous section).

\begin{code}
changeInputs :: D a -> [a] -> D a
changeInputs = WithInputs . graph
\end{code}

A useful metaphor is to think of a neuron graph as a program, and a neuron
diagram as a program combined with its execution for a particular assignment of
inputs.
Because the diagrams \major\ and \both\ share a common neuron graph,
the following evaluates to \cod{True}.

\begin{center}
\begin{figcode}
graph \major == graph \both
\end{figcode}
\end{center}

\noindent
More generally, for any neuron diagram \cod{d} and list of inputs \cod{i}, the
following predicate on \cod{d} and \cod{i} holds.

\begin{center}
\begin{figcode}
graph d == graph (d `changeInputs` i)
\end{figcode}
\end{center}

\begin{figure}[t]
\centering
\includegraphics[scale=\figscale]{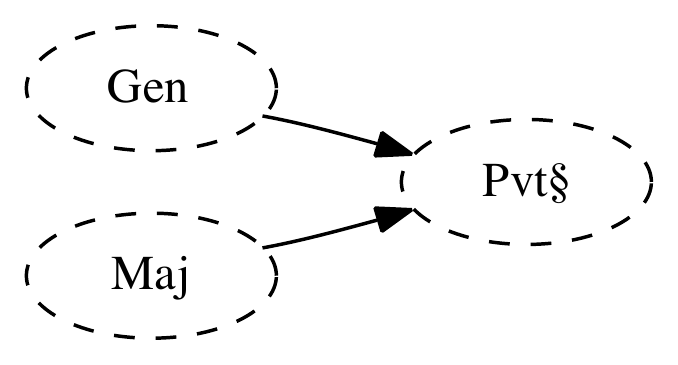}
\caption{The neuron graph underlying both diagrams in
Figure~\ref{fig:orders}.}
\label{fig:ordersG}
\end{figure}

As with diagrams, we can use our DSEL to generate visual representations of neuron
graphs.  We do this with the function \cod{viewG}.
For example, \cod{viewG (graph \major)} produces the neuron graph
shown in Figure~\ref{fig:ordersG}.
We visually distinguish neuron graphs from diagrams by drawing neuron borders
with dashed lines, and because neurons in a graph do not have values (they are
similar to a network of functions), they will never be filled.

There are several other useful DSEL operations (implemented as functions in
Haskell) for querying and manipulating neuron graphs and diagrams.  A small
sample of these are summarized in Figure~\ref{fig:gquery}.
%
%
The \cod{NV} type class in the type of the \cod{allDiagrams} operation will be
defined in Section~\ref{sec:nonbool}, but captures the constraint that
we can only generate all possible diagrams for a neuron graph if the type of
the input values are bounded and enumerable.
One of the more interesting queries is the \cod{asFunction}
operation, which computes the multiple-output function that a graph represents.
That is, given a graph \cod{g} and list of input values \cod{as}, the outputs
of this function are the values of the terminal neurons in the diagram
\cod{D~g~as}.  We call this the \emph{firing semantics} of the graph
\cite{EW10hcc}, but it is important to stress (and we do so repeatedly
throughout this paper) that {\sl the firing semantics of a graph does not
uniquely determine the causal relationships encoded in that graph}.  That is,
the internal structure of a neuron graph is significant; graphs do not simply
reduce to multifunctions.
That said, it is often useful to compare the firing semantics of different
neuron graphs, for example, to confirm that two graphs representing different
causes encode the same effects.  We demonstrate this exact use in the next section,
as part of a larger discussion of modeling stories with neuron diagrams.

\begin{figure}[t]
\centering
\begin{tabular}{ll}
\hline
\cod{neurons~~~~~::~G~a~->~[N~a]}         & all neurons in the graph \\
\cod{asFunction~~::~G~a~->~[a]~->~[a]}    & the multifunction implemented by the graph \\
\cod{graph~~~~~~~::~D~a~->~G~a}       & the graph underlying a neuron diagram \\
\cod{allDiagrams~::~NV~a~=>~G~a~->~[D~a]} & all diagrams generable from the graph \\
\cod{neuronIn~~~~::~Name~->~G~a~->~N~a}   & get a neuron in the graph by name \\
\cod{stateIn~~~~~::~Name~->~D~a~->~a} & the state of a named neuron in the diagram \\
\hline
\end{tabular}
\caption{Basic querying operations on graphs and diagrams.}
\label{fig:gquery}
\end{figure}

%
%
%

\mysection{Basic Causal Modeling}
\label{sec:model}

In this section, we show how a single story can be modeled in different ways,
and how our modeling decisions impact the causal relationships identified in
the story.  An important fact about causal reasoning and neuron diagrams
(stressed above) is that two stories with the same events and the same outcomes
can represent different causes.  Research on causation often revolves around
the problem of finding a particular representation of a story that fits a
preconceived set of causes, and comparing that with other, perhaps more naive
representations.

Returning to our discussion from Section~\ref{sec:intro} of the causal analysis
of the \both\ neuron diagram, recall that the cause of the private's actions
was determined to be a disjunction of the two officers' orders to charge.  This
seems to make sense since either officer's order alone would be sufficient to
make the private charge forward.
But what if charging led the private to run off a cliff and we now find
ourselves in a hearing trying to determine who is at fault for the poor
private's death?  We might still argue that either or both officers are, or we
might argue that only the general, as the highest-ranking officer, is at fault
for this unfortunate outcome.  After all, had the general instead ordered
``retreat!'', the private would have done so despite the major's order to
charge.  
In other words, the causal structure of our neuron graph (underlying both of
the diagrams) does not account for the fact that the general's order, if
given, supersedes the major's.

\begin{figure}[t]
\centering
\includegraphics[scale=\figscale]{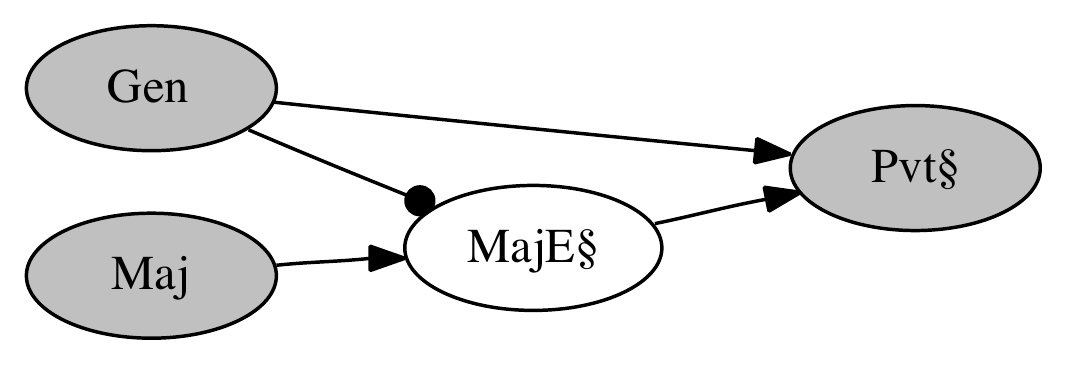}
\caption{Diagram for the scenario where both officers order the private to
charge, but the general's order supersedes the major's order.}
\label{fig:trump}
\end{figure}

As a solution, we introduce an intermediate neuron \MajE\ to represent the
major's ``effective'' order.  If the general gives no order, the major's order
becomes effective, otherwise the general's order prevents this from happening.
To express this notion of trumping prevention, we employ an \emph{inhibitory}
edge that, if the source neuron fires, prevents the target neuron from firing,
regardless or whether or not it is stimulated.  This is represented visually by
a round arrowhead, and represented in the DSEL code by appending
\cod{`InhibBy`} and a list of potentially inhibiting neurons to the end of a
neuron description.
The code defining this new scenario is given below, and the visualized neuron
diagram is shown in Figure~\ref{fig:trump}.

\begin{code}
\trump = diagram [pvt] `WithInputs` [True,True]
  where
    gen  = "Gen"  :# Input
    maj  = "Maj"  :# Input
    majE = "MajE" :# StimBy [maj] `InhibBy` [gen]
    pvt  = "Pvt"  :# StimBy [gen,majE]
\end{code}

\noindent
We see in the visual representation that even though the \Maj\ neuron
stimulated \MajE, this neuron did not fire because it was inhibited by the
firing \Gen\ neuron.  If \Gen\ had not fired, however, then \MajE\ would have
fired as usual.

A causal analysis of our new neuron diagram shows that we have accomplished our
goal.  Although both officers gave the order to charge, the general
alone is determined to be the cause of the private's actions.

\begin{code}
> causes trump
Gen:True ==> Pvt:True
\end{code}

\noindent
The approach that we have taken here, of modeling a story to fit a preconceived
set of causes, is common in causation research \cite{HP05}.  Our DSEL supports
this process by making it easy to create, visualize, and analyze neuron diagrams quickly.
In the next section we show how our extensions to the core language of neuron
diagrams also directly supports this research strategy.
%

Comparing the remodeled story \cod{trump} to the original story \both, we can
see that while the \emph{causal semantics} of the two diagrams differ, their
underlying graphs are \emph{functionally} equivalent.  That is, the private
will charge or not in either model (neuron graph) given the same combination of
inputs.
As mentioned above, we call the mapping from inputs to terminal neuron values
the firing semantics of the neuron graph, and this can be represented as a
multifunction from input values to the values of the terminal neurons of the
graph.
This multifunction can be easily acquired in the DSL with the \cod{asFunction}
operation, but often it is useful to have a more explicit representation of the
firing semantics, for example, so that it can be printed out or compared
directly to the firing semantics of other graphs.  For this, the DSL provides
the \cod{effects} operation, which returns a value of type \cod{Effects~a}.
This name was chosen to reflect that the values of terminal neurons in a neuron
diagram are the subject of causal analysis, that is, they are the effects of
the causes we want to identify.
%
%
We show the effects of our new diagram's underlying graph below.

\begin{code}
> effects (graph \trump)
[Gen:False,Maj:False] -> [Pvt:False]
[Gen:False,Maj:True] -> [Pvt:True]
[Gen:True,Maj:False] -> [Pvt:True]
[Gen:True,Maj:True] -> [Pvt:True]
\end{code}

\noindent
We can confirm that this is identical to the firing semantics of our original graph
by confirming that the following expression evaluates to \cod{True}.

\begin{center}
\begin{figcode}
effects (graph \trump) == effects (graph \both)
\end{figcode}
\end{center}

\noindent
That the firing semantics of these two graphs are equivalent, while the causal
semantics of their corresponding diagrams differ,
%
%
illustrates an important result in causation research: that causal
relationships \emph{cannot} be identified by simply changing the inputs to a
function and observing its outputs.%
\footnote{This is called ``counterfactual reasoning'' in the philosophy
literature.  See Section~\ref{sec:cause}.}
This is the more general form of the point we stressed above. In terms of
neuron diagrams, the multifunction view of a neuron graph is not
sufficient to determine the causal relationships in a story---causation
depends critically not only on the function a graph implements, but
also on its internal structure.

\begin{figure}[t]
\centering
\begin{tabular}{ll}
\hline
\cod{effects~::~NV~a~=>~G~a~->~Effects~a} & the firing semantics of a graph \\
\cod{causes~~::~NV~a~=>~D~a~->~Causes~a}  & the causal semantics of a diagram \\
\hline
\end{tabular}
\caption{Operations for acquiring explicit representations of the two different
semantics.}
\label{fig:semtypes}
\end{figure}

Finally, we provide the types of the \cod{causes} and the \cod{effects}
operations for reference in Figure~\ref{fig:semtypes}.  Note that firing
semantics are a property of neuron graphs, while causal semantics are tied to
an instantiation of a graph with particular input values, that is, a neuron
diagram.
These types again refer to the type class \cod{NV}, which will be defined in
Section~\ref{sec:nonbool}.  Similar to the \cod{Effects~a} type, \cod{Causes~a}
is an explicit representation of a causal semantics that can be pretty printed and
compared to other causal semantics.

In this section we focused on remodeling a problem in order to alter its
inferred causes.
Throughout the discussion we have treated the \cod{causes} operation like an
oracle.  In the next section we demystify it by describing our cause inference
algorithm.
We also introduce our first extension to the language of neuron diagrams, used
to distinguish neurons that are potential causes from those that are not.
The formal definition of our cause inference algorithm can be found in our
previous work \cite{EW10hcc}.



\mysection{Inferring Causes} 
\label{sec:cause}

At the heart of many modern theories of causation is the concept of
\emph{counterfactual reasoning} \cite{Lewis73counter}.  The essence of this
idea is captured in the question: what would have happened if things had been
different?  Given a multifunction, if we change an input and an output also
changes, we say that the output is \emph{counterfactually dependent} on the
input, and so the input is a \emph{cause} of the output.

\begin{figure}[t]
\centering
\includegraphics[scale=\figscale]{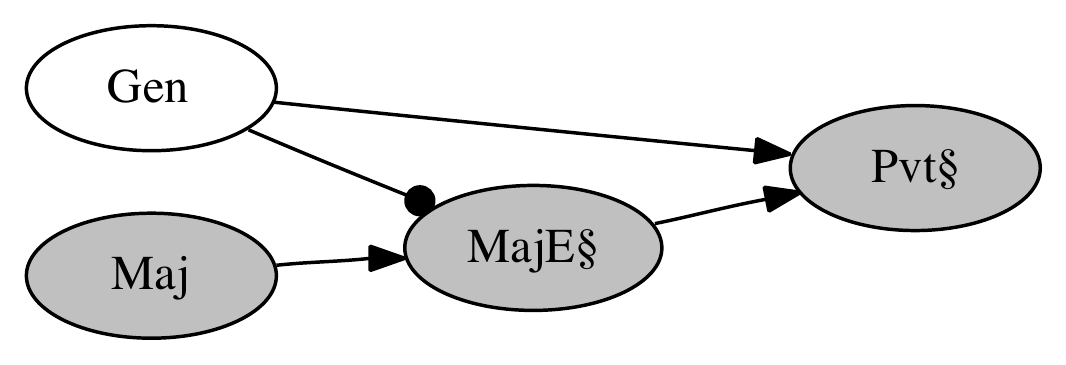}
\caption{Diagram for a variant of the trumping scenario where the general gives no order.}
\label{fig:trumpFT}
\end{figure}

However, as we saw in the previous section, two equivalent multifunctions do not
necessarily produce the same causes.
%
It is easy to construct scenarios that foil direct counterfactual reasoning.
Consider again the diagram in Figure~\ref{fig:trump}, where intuition (and our
\cod{causes} ``oracle'') tells us that the firing of the \Gen\ neuron caused
the firing of the \Pvt\ neuron.
But this cause is not detected by counterfactual reasoning---if we change \Gen\
to not fire (as shown in Figure~\ref{fig:trumpFT}), the value of \Pvt\ does not
change!  The firing of \Maj\ acts as backup and causes \Pvt\ to fire anyway.
This is an example of a classic problem in the philosophy literature known as
\emph{preemption} \cite{Shaeffer00}.

Our cause inference algorithm overcomes this problem, and others that cause
direct counterfactual reasoning to fail, by borrowing from several sources.
The basic idea is to perform counterfactual reasoning locally, overriding the
values of a neuron's immediate predecessors, to determine which neurons form
part of a \emph{causal chain} \cite{Lewis73} back to the ultimate cause of a
neuron's state; then to recursively analyze these neurons.
We will step through an example of this process here.
%
Consider again the neuron diagram in Figure~\ref{fig:trumpFT}, described above
and generated by changing the inputs to the \cod{trump} neuron diagram with the
following DSEL code.

\begin{code}
notTrumped = \trump `changeInputs` [False,True]
\end{code}

\noindent
We begin at the terminal \Pvt\ neuron and examine its two immediate
predecessors, \Gen\ and \MajE.  First we hold all neurons besides \Gen\ and
\Pvt\ fixed and flip the value of \Gen.  We observe that the value of \Pvt\
does not change, so \Gen\ is not recursively analyzed.  We perform the same
test on \MajE\ and observe that, in this case, after changing the value of
\MajE\ the value of \Pvt\ does change, so \Pvt\ is counterfactually dependent
on \MajE.  The \MajE\ neuron is therefore part of a causal chain and so its predecessors
will be recursively analyzed.  If we then test each predecessor of \MajE,
neurons \Gen\ and \Maj, we will find that \MajE\ is counterfactually dependent
on \emph{both} of these neurons.
This implies that both the major's order \emph{and} the general's non-order are
responsible for the private charging.  We can confirm this outcome by
consulting our \cod{causes} oracle.

\begin{code}
> causes notTrumped
Gen:False & Maj:True ==> Pvt:True
\end{code}

%
At first this may seem counterintuitive.  Why isn't the major alone the cause
of the private's actions?
To see the reason for this, suppose we reassign ``real world'' meanings to the
neurons in the diagram.  That is, we leave the structure of the scenario
unchanged, but consider some of the individual neurons to represent different
events.
We will leave the meanings of the \Maj\ and \MajE\ unchanged---they still
represent the major's order to charge and the major's effective order---but now
we will consider the \Gen\ neuron to represent the general's order \emph{to
retreat} (rather than charge) and we will consider the \Pvt\ neuron to
represent carrying out whichever order it receives.
Now when we ask why the private charged, it makes sense to say that the private
charged because the major ordered a charge and the general \emph{did not} order
a retreat.  Had the general ordered a retreat, the \Pvt\ neuron would still
have fired, but the \emph{reason} it fired would have been different, and so
the private's action would be to retreat instead of to charge.
This is just one example of how the causes encoded in a neuron diagram are
often not obvious at first glance or from an informal description of the
scenario it represents.  This demonstrates the value of a formal definition of
the causal semantics of neuron diagrams and an implementation for quickly and
accurately extracting the causal relationships in a diagram.


So far, we have considered counterfactual dependencies between only (causal
chains of) individual neurons.  In fact, the situation is much more
complicated.  Often we need to counterfactually reason about arbitrary boolean
expressions of preceding neurons, and the recursive expansion of these becomes
quite tricky.
The necessity for this more complicated view can be easily seen by prepending
neurons to the above diagram \cod{notTrumped}.  Originally, we identified a
conjunction of neurons as the ultimate cause of the private's action.  If these
neurons have predecessors, however, than this conjunction may not be the
ultimate cause but instead just another step in the causal chain requiring
further expansion.
The details of this reasoning process can be found in \cite{EW10hcc}.


Another potential problem with the causal-chain approach is that we can recurse
too far.  As the algorithm has been described, the implicit base case is an
exogenous neuron.  But we can almost always arbitrarily prepend neurons
without significantly altering the story.  For example, we could add a
neuron \Wake\ that stimulates the \Gen\ neuron in the diagram in
Figure~\ref{fig:trump}
and represents the general waking up that morning.  After all, the general
cannot order the private to charge if asleep!  Now we will identify the firing
of \Wake\ as the cause of the private charging, which seems odd.
While this example is kind of silly, the underlying question of the
transitivity (or not) of causation is a hotly debated topic among philosophers
\cite{Mackie80,Hall00,Hitchcock01}.

\begin{figure}[t]
\centering
\includegraphics[scale=\figscale]{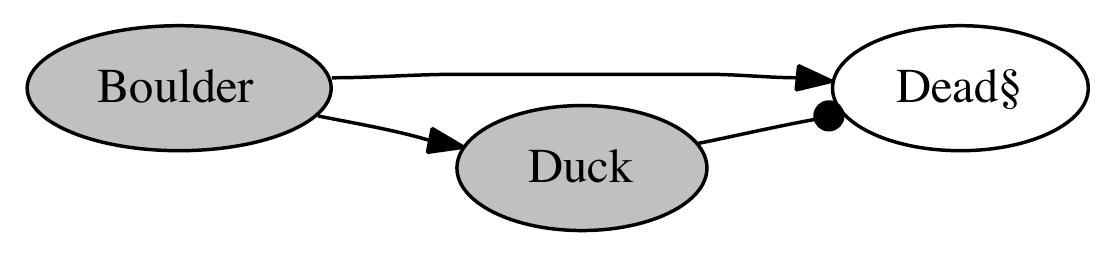}
\caption{Diagram for the scenario in which a hiker avoids a falling boulder by ducking.}
\label{fig:boulder}
\end{figure}

In fact, assuming that causation is strictly transitive can lead to
paradoxes.  This is demonstrated by the example in Figure~\ref{fig:boulder}, in
which a boulder falls down a hill toward a hiker who subsequently ducks and
therefore does not die.
If we perform our naively recursive causal chain analysis, we identify \Duck\
as the first link in the chain, followed by \Boulder.  So the very boulder
which threatened the hiker's life is identified as the cause of the hiker's
survival!

In \cite{EW10hcc} we adapt a pragmatic solution to this problem originally
developed in another language \cite{HP05}.  We simply explicitly distinguish
neurons that are potential causes, called \emph{actions}, from those that are
not, called \emph{laws}.
We then modify our algorithm to stop recursing whenever an action is
encountered.
In the visual representation of neuron diagrams, we annotate laws with a
``\lawSym'' symbol and leave actions unadorned.
Intuitively, actions represent points in a story where things \emph{could} have
gone differently (for example, decision points by the actors in the scenario),
whereas laws represent hard-wired relationships or parts of the story that are
simply accepted as given.
%
This is an essential causal modeling feature previously absent from
neuron diagrams.


When creating neuron diagrams in our DSEL, by default, input neurons are
actions and all other neurons are laws.  This works most of the time, and is
nearly equivalent to the naively recursive algorithm.%
\footnote{The only exceptions to this equivalence are exogenous, non-input, law
neurons, such as the constant-valued neurons introduced in
Section~\ref{sec:ndinst}.}
When we need to override this behavior, however, we can extend a neuron
description with an explicit \cod{IsKind} annotation.  This is demonstrated in
the definition of the \Duck\ neuron in the following definition of the boulder
problem.

\begin{code}
boulder = diagram [dead] `WithInputs` [True]
  where
    boulder = "Boulder" :# Input
    duck = "Duck" :# StimBy [boulder] `IsKind` \Act
    dead = "Dead" :# StimBy [boulder] `InhibBy` [duck]
\end{code}

\noindent
Because \Duck\ is identified as an action, the algorithm will halt and return
\Duck\ as the cause of the hiker's survival rather than recursively analyzing
its predecessors.  We show the result of the causal analysis below.

\begin{code}
> causes boulder
Duck:True ==> Dead:False
\end{code}

We have now seen several basic applications of our DSEL. 
Before we move on to more advanced applications, including defining our own
types of neurons and creating diagrams that operate on non-boolean values, we
return to the topic of neuron representation from Section~\ref{sec:basics}.

\mysection{Neuron Descriptions}
\label{sec:rep}

So far, we have considered only the concrete syntax of neuron descriptions.
This has included the definition of input neurons, standard neurons with
stimulating and inhibiting edges, and annotations for overriding the default
kind of a neuron.
As mentioned in Section~\ref{sec:basics}, this notation for describing neurons
can be considered a mini-DSL within the larger DSL for creating neuron
diagrams.  Because users of neuron diagrams often invent new types of neurons
on-demand, this mini-DSL must be very extensible.  In this section we show how
our design directly supports this requirement.

One of the major advantages of DSELs is that their concrete syntax is
inherently extensible; one can always add new constructs to the language by
simply defining new functions.  Extensibility can still be limited, however, by
the underlying representation of the semantics and abstract syntax of the
language, which are usually captured in types.  For example, we could represent
the entire syntax of our neuron description mini-DSL so far in a single data type,
but this is not very extensible---new constructs would be limited to mere
syntactic sugar and adding new types of neurons would be impossible.


Instead, we represent neuron descriptions as an open class of types.  Through
local quantification of the description type parameter in the type \cod{N~a}
(repeated below), we enable the use of different description types in a single
neuron diagram.  This is a highly extensible representation, that allows users
to define their own neuron description types and use them interchangeably with
those provided by the DSEL.

In Section~\ref{sec:ndclass} we present the neuron description type class
\cod{ND} and its related types. 
We show how to define basic neuron descriptions in Section~\ref{sec:ndinst},
and more complex instances that modify or compose existing descriptions in
Section~\ref{sec:dec}.
Finally, we briefly consider an alternative representation of neuron
descriptions in Section~\ref{sec:ndexp}, and argue that the chosen approach
best supports the design goals of simplicity and extensibility.

\mysubsection{The Neuron Description Type Class}
\label{sec:ndclass}

A neuron description must provide several pieces of information: (1) the kind
of the neuron (action or law); (2) the function that the neuron implements,
called its \emph{firing function}; (3) the visual style of the neuron; and (4)
the incoming edges from preceding neurons, which each have their own visual
style.
We will consider the representation of each of these components separately, but
first, we repeat the definition of the neuron type \cod{N~a} below for
reference.

\begin{code}
data N a = forall d. ND d a => Name :# d a
\end{code}

\noindent
Again note that from this definition, we see that \cod{ND} is a multi-parameter
type class on a type constructor \cod{d}, representing the description type,
and the neuron value type \cod{a}.

First, we represent the kind of a neuron, motivated in the previous section,
with the following straightforward data type \cod{Kind}.

\begin{code}
data Kind = \Act | Law
\end{code}

Second, the firing function of a neuron is a function from a list of input
values (the values of its predecessor nodes) to a result value, that is,
\cod{[a]~->~a}.  However, the input neurons of a diagram do not have a firing
function, and we represent this optionality with a \cod{Maybe} type.

\begin{code}
type Fire a = Maybe ([a] -> a)
\end{code}

\noindent
Therefore, an input neuron, such as \Gen\ in the previous examples, has a
firing function of \cod{Nothing}.
The implementation of a non-input neuron's firing function usually depends on
the type of values processed by the neuron.  For example, a standard boolean
neuron with only stimulating edges, such as the \Pvt\ neuron in our examples,
would have a firing function of \cod{Just or}, where \cod{or} is the standard
Haskell function for disjunction of a boolean list.
It is this dependency of the firing function on the value type that motivates
the definition of \cod{ND} as a multi-parameter type class.  Often we will want
to define neurons that are stylistically and structurally identical, but which
operate on different value types.  This type of extensibility is easily
supported by this construction, and we will see an example of redefining
standard neurons with a new value type in Section~\ref{sec:nonbool}.

Third, the visual style of a neuron---essential for determining the function a
neuron implements in the visual representation---is represented by a list of
GraphViz attributes (name-value pairs of strings), and captured in the type
\cod{Style}.
%
%
%
A suite of functions is provided for creating the most commonly used
attributes and styles.  This not introduces a
layer of abstraction that increases modularity and minimizes the dependency on
GraphViz.
By hiding the definition of the \cod{Style} type and exposing only the
functional interface we can add new visualization back-ends without breaking
existing DSEL code.

Finally, each incoming edge to a neuron also has its own style, which is also
significant to determining the firing function of a neuron from its visual
representation.  For example, we indicate stimulating edges with triangular
arrowheads and inhibiting edges with circular arrowheads.  We thus represent an
edge as a pair of a source neuron and a style.

\begin{code}
type Edge a = (N a, Style)
\end{code}

\noindent
Note that the destination neuron of an edge is implicitly the neuron containing
the description that contains that edge, and so is not represented explicitly
in the edge type.

Taking all of the above, we can define the neuron description type class,
\cod{ND} as follows.

\begin{code}
class ND d a where
  kind  :: d a -> Kind
  fire  :: d a -> Fire a
  style :: d a -> Style
  edges :: d a -> [Edge a]
  kind  _ = Law
  style _ = []
  edges _ = []
\end{code}

\noindent
Useful defaults are provided for all of the methods in the type class except
for \cod{fire}, so it is possible for some definitions to be quite small, as we
will see in the next subsection.

\mysubsection{Basic Neuron Descriptions}
\label{sec:ndinst}

\begin{figure}[t]
\hfill
\begin{figcode}
data Input a = Input
instance ND Input a where\ \ \ \ \ \\
  kind _ = Action
  fire _ = Nothing
\end{figcode}
\hfill\vline\hfill
\begin{figcode}
data Const a = Const a
instance ND Const a where
  fire (Const a) = Just (const a)\ \ \ \ \ \ \ \ \ \ \\

\end{figcode}
\hfill{}
\caption{Neuron descriptions for basic exogenous neurons.}
\vspace{-1ex}
\label{fig:exos}
\end{figure}

In this subsection we show the definitions of four basic neuron description
types.  We begin with the definition of two descriptions for exogenous neurons,
shown in Figure~\ref{fig:exos}. On the left is the \cod{Input} neuron
description that is used throughout the paper.
This description is represented by just a nullary constructor, but note that we
must make \cod{Input} a type constructor to satisfy the constraints of the
\cod{ND} type class.%
\footnote{An unreferenced type parameter like \cod{a} is sometimes called a
\emph{phantom type} \url{http://www.haskell.org/haskellwiki/Phantom_type}. This
type information could be used, for example, to alter the style of input
neurons of different types.}
Input neurons are usually actions, so in the type class instance we override
the default \cod{kind} method.  Input neurons also have no firing functions, so
we set \cod{fire~Input} to \cod{Nothing}.
On the right, we define a simple description for constant-valued neurons.
While these neurons are exogenous like inputs, they are not actions by default,
so we rely on the default \cod{kind} method in the instance declaration.

\begin{figure}[t]
\hfill
\begin{figcode}
data StimBy a = StimBy [N a]
instance ND StimBy Bool where
  fire  _           = Just or
  edges (StimBy ns) = plain ns 

\end{figcode}
\hfill\vline\hfill
\begin{figcode}
data Thick a = Thick Int [N a]
instance ND Thick Bool where
  fire  (Thick k _)  = Just ((>=k) . count)
  style _            = penwidth 3
  edges (Thick _ ns) = plain ns
\end{figcode}
\hfill{}
\caption{Neuron descriptions for basic boolean endogenous neurons.}
\label{fig:endos}
\vspace{-1ex}
\end{figure}

In Figure~\ref{fig:endos} we define a couple of basic descriptions for
endogenous, boolean neurons.  While we left the value type parameter unfixed in
the instance declarations for \cod{Input} and \cod{Const} (enabling these
descriptions for use with all value types), in these descriptions we fix the
type in the instances to \cod{Bool} since the firing functions are specific to
boolean values.
%
%
The \cod{plain} function used in these definitions takes a list of predecessor
neurons and returns a list of edges with ``plain'' styles, that is, with
standard line thickness and triangular arrowheads.
On the left, we define the \cod{StimBy} neuron description that has been used
throughout the paper. Its firing function is just a disjunction of its inputs.
On the right, we define thick-bordered neurons that, given a parameter \cod{k},
fire if they are stimulated by at least \cod{k} predecessors. The expression
\cod{penwidth 3} produces a thick-bordered style
while \cod{count} is a function that returns the number of \cod{True} values in
a list.
%
%
%
Example uses of thick neurons will be given in Section~\ref{sec:expl}.

%
%

The DSEL provides several other neuron descriptions for standard neurons types
used in the philosophy literature.  These include diamond-shaped XOR neurons,
that fire if they are stimulated by exactly one predecessor, and neurons with
\emph{un}stimulating edges that fire if at least one of their predecessors did
not fire. The definition of this latter neuron description type,
\cod{UnstimBy}, is similar to the definition of \cod{StimBy}, except that
unstimulating edges are represented visually by a hollow arrowhead. This
neuron description will be used in Section~\ref{sec:logic}.
In the next section we provide examples of more complex neuron descriptions
that modify and combine other descriptions.

\mysubsection{Description Decorators and Composition}
\label{sec:dec}

In addition to the core neuron descriptions described above, we have also seen
two examples of annotation-like constructs in the mini-DSL for neuron
descriptions.  The first, \cod{`InhibBy`}, is used to add inhibiting edges to a
neuron, while the second, \cod{`IsKind`}, is used to set the kind of a neuron
explicitly.
Knowledge of Haskell syntax and typing reveals that these annotations are
implemented as data constructors applied as infix operators, and that the
resulting data type value must be an instance of the neuron description type
class.
These description constructors take another neuron description as their first
argument, ``wrapping'' them and extending or tweaking their functionality.  The
implementation of these descriptions is very similar to a well-known idiom in
the object-oriented programming community called the \emph{decorator} design
pattern \cite{Gamma1995}.

First, we examine the \cod{`IsKind`} annotation, the simpler of the two
built-in decorators seen so far.  We define the data type for this construct as
follows; its constructor (the \cod{IsKind} keyword in the DSL) accepts a neuron
description and a \cod{Kind} value as arguments.

\begin{code}
data IsKind d a = IsKind (d a) Kind
\end{code}

\noindent
The \cod{ND} instance for this type requires that the wrapped description
\cod{d} also instantiates the type class for the given value type \cod{a}.  It
then simply defers to the wrapped description in all cases except for the
\cod{kind} method, which is overridden to the argument value.

\begin{code}
instance ND d a => ND (IsKind d) a where
  kind  (IsKind _ k) = k
  fire  (IsKind d _) = fire d
  style (IsKind d _) = style d
  edges (IsKind d _) = edges d
\end{code}

\noindent
This demonstrates a very flexible and powerful way to extend the syntax and
functionality of neuron descriptions.  The \cod{IsKind} decorator can now be
applied to \emph{any} neuron description to set its kind explicitly.

Although slightly more complicated than the above, the decorator pattern can
also be used to add new types of edges to the existing language of neuron
descriptions.  We demonstrate this by implementing the \cod{`InhibBy`}
construct for adding inhibiting edges to any neuron description.
First we define the data type representing this construct, as before.  The
first argument is a neuron description to wrap, and the second is a list of
inhibiting predecessor neurons.

\begin{code}
data InhibBy d a = InhibBy (d a) [N a]
\end{code}

\noindent
In the \cod{ND} instance for this construct, we will defer to the wrapped
description's \cod{kind} and \cod{style} methods and extend the wrapped
description's \cod{fire} and \cod{edges} methods as described below.

\begin{code}
instance ND d Bool => ND (InhibBy d) Bool where
  kind  (InhibBy d _)  = kind d
  style (InhibBy d _)  = style d
  fire  (InhibBy d ns) = extend d (&&) (all not)
  edges (InhibBy d ns) = edges d ++ styled (arrowhead "dot") ns
\end{code}

\noindent
Extending the list of edges is straightforward---we just concatenate the new
edges to the end of the wrapped description's edges, adding a \cod{Style} value
to each that will draw the new edges with a circular arrowhead.  
The
\cod{arrowhead} function produces this style, and the \cod{styled} function
applies a style to every neuron in a list, producing a list of edges.
To extend the firing function, we rely on a helper function \cod{extend}.  The
type of this function is given below.

\begin{code}
extend :: ND d a => d a -> (a -> b -> a) -> ([a] -> b) -> Fire a
\end{code}

\noindent
This function extends the wrapped description's firing function by evaluating
the original firing function on the original predecessors, applying the
function passed as the third argument to the new predecessors, and combining
these results with the function passed as the second argument.  In this
example, we apply \cod{all not} to the inhibiting predecessors and combine the
result with the original firing function with the {\ttsize\verb#(&&)#} function.  That
is, the resulting neuron will fire if the wrapped description indicates that it
should fire, and none of the inhibiting neurons fire.

Note that while the \cod{IsKind} decorator can be applied to any description
regardless of the value type \cod{a}, the \cod{InhibBy} decorator is limited to
operating within boolean valued neuron diagrams.  This is because the firing
function is defined in terms of boolean functions (\cod{not} and
{\ttsize\verb#(&&)#}),
and the meaning of an inhibiting edge is not clear in arbitrary non-boolean
domains.
One of the strengths of this language representation, however, is that existing
neuron descriptions on specific value types can be easily extended to new value
types by simply re-instantiating the \cod{ND} type class with a different
second type argument.  This is demonstrated on the \cod{InhibBy} type in
Section~\ref{sec:nonbool}.

The decorator pattern described above is easy to follow and provides a flexible
form of extensibility.  However, it also enforces a hierarchy on neuron
descriptions that is sometimes arbitrary.  Some descriptions are ``cores'',
like \cod{StimBy}, while others are decorators, like \cod{InhibBy}.  In the
case of stimulating and inhibiting edges, this hierarchy follows
convention---inhibiting edges are added to all sorts of boolean neurons
(including, for example, the XOR and thick neurons described as the end of the
previous subsection), while stimulating edges that implement logical
disjunction are not.
When a natural hierarchy does not exist, however, or when we want to break with
convention, the design also supports more ad hoc and symmetric forms of
composition.
We provide two such composition constructs for boolean neuron descriptions
below.
The {\ttsize\verb#:&&:#} construct combines two descriptions by merging their
styles, concatenating their edges, and combining their firing functions with
the Haskell {\ttsize\verb#(&&)#} function.  The {\ttsize\verb#:||:#} construct is similar, except
that it combines its arguments' firing functions with {\ttsize\verb#(||)#}.

\begin{code}
data And l r a = l a :&&: r a
data Or  l r a = l a :||: r a
\end{code}

\noindent
We show the \cod{ND} instance for the {\ttsize\verb#:||:#} construct below.  The
{\ttsize\verb#:&&:#} instance is identical, except that it uses
{\ttsize\verb#(&&)#} as the
second argument to \cod{extend}.

\begin{code}
instance (ND l Bool, ND r Bool) => ND (Or l r) Bool where
  kind  (_ :||: r) = kind r
  style (l :||: r) = style l ++ style r
  edges (l :||: r) = edges l ++ edges r
  fire  (l :||: r) = extend l (||) (fromJust (fire r))
\end{code}

\noindent
Note that this definition is not \emph{purely} symmetric---the kind of the
right argument is explicitly preferred and the style of the right argument will
also be preferred in the case of clashes, though this is an implicit property
of the \cod{Style} type.

These constructs provide another example of how the chosen representation
fulfills our design goals of extensibility and flexibility, essential qualities
for supporting neuron diagram use in practice.
However, the type class-based approach is not the only way to achieve these
design goals, and is arguably heavier weight than some of the alternatives.  In
the next section we compare our approach with a more explicit representation of
neuron descriptions, and argue in favor of our choice of representation.

\mysubsection{Comparison to a Direct Data Type Representation}
\label{sec:ndexp}

\newcommand{\NDExp}{ND'}

An alternative to the type class-based implementation of neuron
descriptions, is a more direct representation
where the \cod{ND} type class is replaced by a data type that
contains values corresponding to each of the four methods in \cod{ND}.
We will call this data type \cod{\NDExp} to distinguish it from the
\cod{ND} type class.

\begin{code}
data \NDExp a = \NDExp Kind (Fire a) Style [Edge a]
\end{code}

\noindent
This representation is nearly as extensible as the type class-based approach
and depends only on unextended
Haskell~98~\cite{Haskell98}.  Core neuron descriptions can be defined as
functions that produce values of type \cod{\NDExp} and decorators can be
defined simply as functions that accept values of this type as arguments and
produce them as results.  We can do all of this in a way that changes the
concrete syntax of neuron descriptions very little, simply replacing the
capitalized names of Haskell data constructors with the lowercase names of
functions.
The neuron description definitions are also often terser in the direct
representation since we do not have the extra syntactic overhead of
declaring a data type and instantiating a type class.
So, if the representations are nearly equivalent, why do we use the more
verbose alternative?

We choose the current approach over the data type representation for three
reasons: (1) it fundamentally supports the extension of existing neuron
descriptions to new value types by instantiating the multi-parameter \cod{ND} type class
with new value types, promoting reuse of the syntax and structure
of existing neurons; (2) it allows us to explicitly manipulate descriptions
and encode constraints between neuron descriptions in the type system; and (3)
the type class approach seamlessly integrates the data type approach, but not
vice versa.
The first point is illustrated in the next section, where we extend the
\cod{StimBy} and \cod{InhibBy} descriptions to a non-boolean domain.  We
briefly demonstrate the other two points below.

The abilities to explicitly manipulate neuron descriptions and to use Haskell
types to encode syntactic constraints in the neuron description DSL are both
demonstrated by the following syntactic extension to the description DSL that
adds a single stimulating neuron to an existing \cod{StimBy} description.

\begin{code}
addStim :: StimBy a -> N a -> StimBy a
addStim (StimBy ns) n = StimBy (ns ++ [n])
\end{code}

\noindent
With this extension, we can write, for example, a description of a neuron
stimulated by predecessors \cod{a}, \cod{b}, and \cod{c} as
{\ttsize\verb#StimBy [a,b] `addStim` c#}.
An important fact about this new construct is that it can \emph{only} be
applied to a \cod{StimBy} description.  It is impossible to define an extension
with the same constraint in the data type representation since the
\cod{StimBy} description is never represented explicitly---it would instead be
implemented as a function that produces a generic value of type \cod{\NDExp}.
While this example is somewhat contrived, it demonstrates a fundamental
advantage in expressiveness for the type class-based representation, and one
that is likely to become increasingly useful as extensions to the language
grow more complex.

The third reason that we prefer the type class-based representation is that
it seamlessly integrates the direct data type representation, while the
converse is not true.  This means that users preferring the direct
representation can use it freely, even using both representations together in
the same diagram.
To integrate the direct representation into the type class-based approach, we add
the following trivial instance of the \cod{ND} type class for the \cod{\NDExp}
data type.

\begin{code}
instance ND \NDExp a where
  \OBRACE kind  (\NDExp k _ _ _) = k ; style (\NDExp _ _ s _) = s ;
    fire  (\NDExp _ f _ _) = f ; edges (\NDExp _ _ _ e) = e \CBRACE
\end{code}


\noindent
It is also easy to convert a type class-based neuron description into an
explicit neuron description, for example, using the following function
\cod{to\NDExp}.

\begin{code}
to\NDExp :: ND d a => d a -> \NDExp a
to\NDExp d = \NDExp (kind d) (fire d) (style d) (edges d)
\end{code}

\noindent
However, the interoperability is not as seamless in this direction since it
requires applications of \cod{to\NDExp} to be sprinkled throughout DSEL code.

\mysection{Beyond Boolean Causation}
\label{sec:nonbool}

Although we have considered only boolean neuron values so far, an important
contribution of this work is the extension of neuron diagrams to non-boolean
values.
In order to perform counterfactual reasoning on non-boolean values, we must be
able to enumerate all possible values that a neuron can take on.  We thus
require that a neuron value type is bounded and enumerable.  We also must be
able to distinguish different values visually, and we capture these visual
properties in the \cod{Style} type described in Section~\ref{sec:ndclass}.  We
express all of these requirements in the following type class \cod{NV} for
neuron values.

\begin{code}
class (Bounded a, Enum a, Eq a) => NV a where
  valStyle :: a -> Style
\end{code}

\noindent
We instantiate this type class for boolean values as follows, where
\cod{fillWith} returns a style that fills the node with the argument color.

\begin{code}
instance NV Bool where
  valStyle True  = fillWith "gray"
  valStyle False = []
\end{code}

In the rest of this section we extend our running general-major-private example
to non-boolean values.
Recall the diagram \cod{trump}, shown in
Figure~\ref{fig:trump}, where the general's order to charge trumps the major's.
In Section~\ref{sec:cause} we briefly considered a variation of this story in
which the general orders a retreat.  What if instead we allowed \emph{both}
officers in this scenario to order \emph{either} a charge or a retreat, or to
issue no order at all?
We capture these three possibilities in the following bounded, enumerable data
type.

\begin{code}
data Order = None | Charge | Retreat
\end{code}

\noindent
To use this data type in visualized neuron diagrams, we also
instantiate the \cod{NV} type class, setting \cod{Charge} and \cod{Retreat} to
be colored green and red, respectively, and leaving neurons with \cod{None}
values unfilled.\footnote{Color names use the X11 naming scheme
\url{http://www.graphviz.org/doc/info/colors.html\#x11}.}

\begin{code}
instance NV Order where
  valStyle None    = []
  valStyle Charge  = fillWith "palegreen"
  valStyle Retreat = fillWith "orangered"
\end{code}

Now, we would expect that if the general gives an order, the private will carry
out that order, otherwise the private will carry out the order (or non-order)
given by the major.  There are at least two ways to encode these firing
semantics in a neuron graph.  The first is to extend the notion of stimulating
and inhibiting edges to the \cod{Order} data type, then reuse the neuron graph
from the diagrams in Figure~\ref{fig:trump} and Figure~\ref{fig:trumpFT}.
The second is to invent a new type of neuron that interprets orders from
multiple officers, taking into account the officers' ranks, thereby
encapsulating all of the logic in this new neuron's firing function.
We demonstrate both approaches here, and show that each can be used to create
graphs with the same firing semantics, and corresponding diagrams with the same
causal semantics.

%
%

We first consider the approach of extending the existing neuron descriptions,
for adding stimulating and inhibiting edges, to work with \cod{Order} values.
We must begin by asking what it means to extend the concepts of stimulating and
inhibiting edges to this new data type.
In this case, it seems that a \cod{None} value corresponds to a non-firing
neuron in the boolean representation, while \cod{Charge} and \cod{Retreat}
values correspond to firing neurons.  Therefore, if a neuron \cod{n} is
stimulated only by predecessors \cod{ps} with \cod{None} values, then \cod{n}
should also have a value of \cod{None}; if at least one of \cod{ps} has a
\cod{Charge} or \cod{Retreat} value, however, then \cod{n} should have the same
value.  
It is not immediately clear what to do if some neurons in \cod{ps} have
\cod{Charge} values and some have \cod{Retreat} values.  One possibility is to
simply default to \cod{None} in this case, while another is to set \cod{n} to
the value that is higher represented in \cod{ps} (and to \cod{None} if
\cod{Charge} and \cod{Retreat} are represented equally); that is, if two
neurons in \cod{ps} have the value \cod{Charge} and one has the value
\cod{Retreat}, we set \cod{n} to \cod{Charge}.
We choose the second approach here, and this logic is captured in the following
helper function \cod{resolve}.

\begin{code}
resolve :: [Order] -> Order
resolve os | c > r     = Charge
           | r > c     = Retreat
           | otherwise = None
  where [c,r] = [length (filter (==o) os) | o <- [Charge,Retreat]]
\end{code}

\noindent
Using this function, we can instantiate the \cod{ND} type class for stimulated
neurons with \cod{Order} values by simply copying the corresponding instance
for \cod{Bool} (from Section~\ref{sec:ndinst}) and replacing \cod{or} in the
firing function with \cod{resolve}.

\begin{code}
instance ND StimBy Order where
  fire  _ = Just resolve
  edges (StimBy ns) = plain ns 
\end{code}

\noindent
With this extension, we can create a non-boolean variant of the stories
represented by the neuron diagrams in Figure~\ref{fig:orders}.  In this
variant, the private carries out the order given to him by either officer, but
gets confused if the officers give conflicting orders, and responds by doing
nothing.

The case for inhibiting edges is similar in that we separate orders into cases
corresponding to boolean firing neurons (\cod{Charge} and \cod{Retreat}) and
non-firing neurons (\cod{None}).  That is, if any inhibiting predecessor of a
neuron \cod{n} has a non-\cod{None} value, we override the value of \cod{n} to
be \cod{None}.
We implement this again by copying and modifying the corresponding \cod{ND}
instance for \cod{InhibBy} on boolean values (from Section~\ref{sec:dec}).
Only the firing function is different from the boolean case, so we present only
the \cod{fire} method below, replacing the rest with an ellipsis.

\begin{code}
instance ND d Order => ND (InhibBy d) Order where
  fire (InhibBy d ns) = extend d (\BSLASH{o} b -> if b then None else o) (any (/=None))
  \ldots
\end{code}

\noindent
As in the boolean case, we rely on the \cod{extend} helper function for
extending the decorated neuron description's firing function.  On the
inhibiting predecessors (those bound by the \cod{InhibBy} decorator), we apply
the function \cod{any~(/=None)}, which returns \cod{True} if the neuron's
return value should be overridden with \cod{None}.
The function passed as the second argument to \cod{extend} combines with
boolean value with the order returned by the decorated description's firing
function, implementing the overriding behavior.

One of the major advantages of the type class-based representation of neuron
descriptions, discussed in Section~\ref{sec:rep}, is that we can easily extend
existing neuron description types to work with new value types, as we have done here
with \cod{StimBy} and \cod{InhibBy}.  This means that we can reuse the graph of
the boolean version of the \cod{trump} neuron diagram directly in both boolean
and non-boolean versions of the story.
To do this, we define the graph independently as follows.  Note that this is
exactly the same graph structure as in the definition of the \cod{trump} diagram.

\begin{code}
trumpG = Graph [pvt]
  where
    gen  = "Gen"  :# Input
    maj  = "Maj"  :# Input
    majE = "MajE" :# StimBy [maj] `InhibBy` [gen]
    pvt  = "Pvt"  :# StimBy [gen,majE]
\end{code}

\noindent
We can then create both boolean and non-boolean diagrams from this graph by
simply instantiating it with different inputs, as shown below.

\begin{code}
trumpBool  = trumpG `WithInputs` [True,True]
trumpOrder = trumpG `WithInputs` [Charge,Retreat]
\end{code}

\begin{figure}[t]
\centering
\subcaptionbox{\label{fig:retreat:CR}}
  {\includegraphics[scale=\figscale]{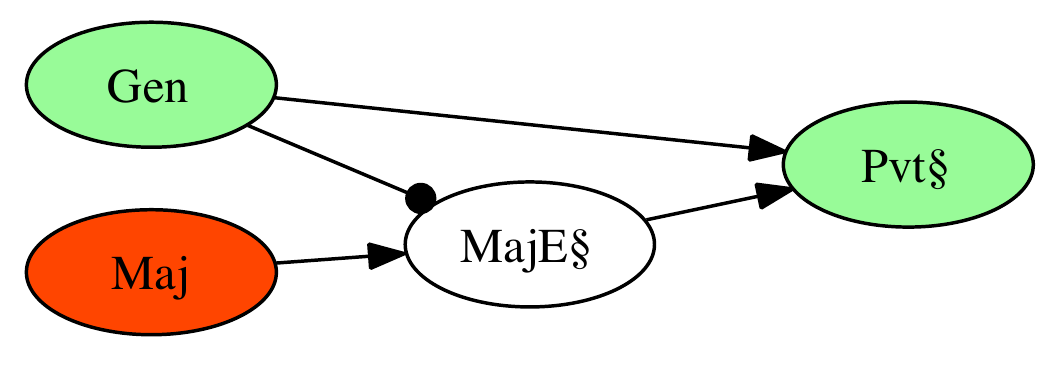}}
\quad
\subcaptionbox{\label{fig:retreat:NR}}
  {\includegraphics[scale=\figscale]{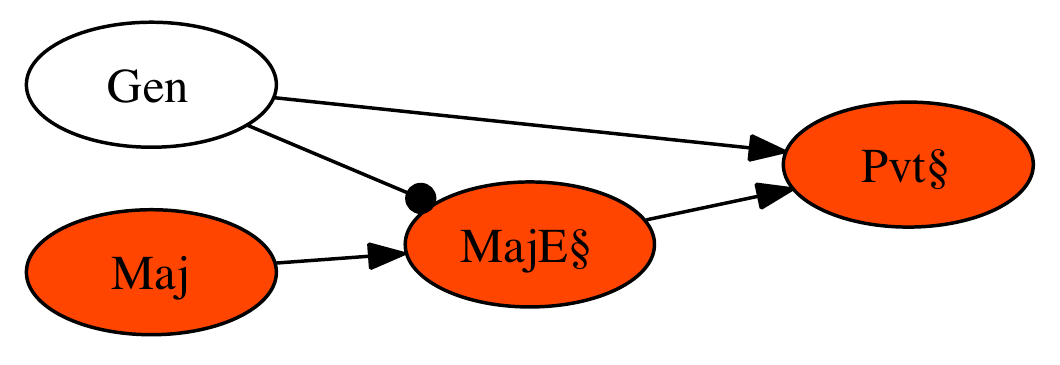}}
\caption[]{Neuron diagrams for a private reacting to orders from two officers.
In \sref*{fig:retreat}{CR} the general orders the private to charge while the major orders
a retreat, so the private charges.  In \sref*{fig:retreat}{NR} the major orders
the private to retreat and the general is silent, so the private retreats.}
\label{fig:retreat}
\end{figure}

\noindent
The \cod{trumpBool} diagram is exactly equivalent to \cod{trump}, shown in
Figure~\ref{fig:trump}, where both officers give the order to charge but it is
the general's order that is carried out.
The trumping aspect of this graph is made more explicit in the non-boolean
case.  In the \cod{trumpOrder} diagram, shown in Figure~\sref{fig:retreat}{CR},
the general orders a charge (colored green, or light gray in black and white)
while the major orders a retreat (red, darker gray).  We can see clearly in
the diagram that it is the general's order that takes precedence since the
\Pvt\ neuron is colored green, indicating that the private charged.
If the general does not issue an order, then the private carries out the
major's order, as shown in Figure~\sref{fig:retreat}{NR}, which was created
with the DSEL code {\ttsize\verb#trumpG `WithInputs` [None,Retreat]#}.

%
%

Next we consider the second approach to extending the general-major-private
example to the non-boolean \cod{Order} type.  This time we will extend the
syntax with a new neuron description that resolves orders according to the rank
of the issuing officers.
We call our new neuron description \cod{ByRank}, and we assume that the
officers that are its predecessors are sorted in decreasing order of rank.  We
can represent this neuron description with the following simple data type.

\begin{code}
data ByRank a = ByRank [N a] 
\end{code}

\noindent
To make \cod{ByRank} a neuron description, we must instantiate the \cod{ND}
type class.  We give the neuron a unique shape in order to visually distinguish
it from other neurons, and define the firing function to simply return the
first non-\cod{None} order received by one of its predecessors (or \cod{None}
if no such order is found).  This will be the order by the highest ranking
officer since the predecessors of a \cod{ByRank} neuron are sorted.

\begin{code}
instance ND ByRank Order where
  fire  _ = Just (maybe None id . find (/=None))
  style _ = shape "pentagon"
  edges (ByRank ns) = plain ns 
\end{code}

\noindent
Now we can redefine the story represented by the \cod{trumpOrder} diagram in
terms of our new neuron description as follows.

\begin{code}
byRank = diagram [pvt] `WithInputs` [Charge,Retreat]
  where 
    gen = "Gen" :# Input
    maj = "Maj" :# Input
    pvt = "Pvt" :# ByRank [gen,maj]
\end{code}

\begin{figure}[t]
\centering
\subcaptionbox{\label{fig:process:CR}}
  {\includegraphics[scale=\figscale]{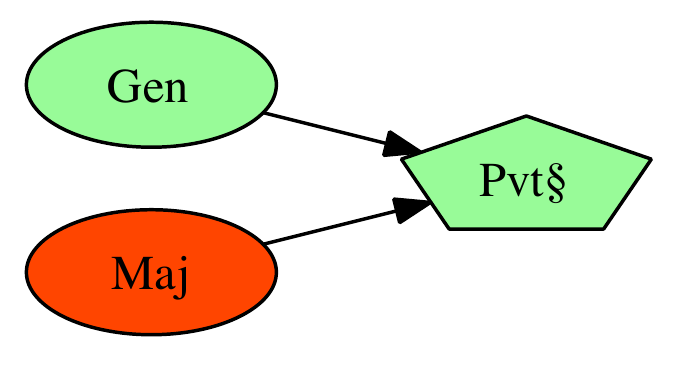}}
\qquad
\subcaptionbox{\label{fig:process:NR}}
  {\includegraphics[scale=\figscale]{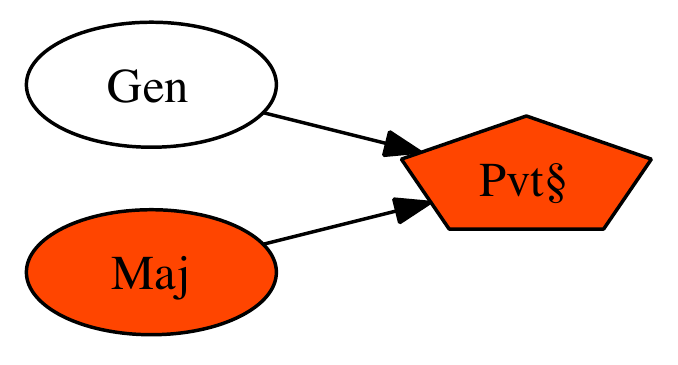}}
\caption[]{Neuron diagrams for a private processing orders from two officers.
In \sref*{fig:process}{CR} the general orders the private to charge while the major orders
a retreat, so the private charges.  In \sref*{fig:process}{NR} the major orders
the private to retreat and the general is silent, so the private retreats.}
\label{fig:process}
\end{figure}

\noindent
This diagram is shown in Figure~\ref{fig:process}, along with the variant
diagram in which the major issues no order, created with the DSEL code
{\ttsize\verb#byRank `changeInputs` [None,Retreat]#}.

%
%

An interesting feature of the graphs we have created with these two different
approaches is that, not only do they have the same firing semantics, they also
encode the same causes.  We can easily check this claim with the DSEL.
First, we check that the firing semantics of the two graphs are the same by
comparing their effects.

\begin{code}
> effects trumpG == effects (graph byRank)
True
\end{code}

\noindent
Next, we want to confirm that the graphs also encode the same causal
relationships.  Since the causal semantics are defined in terms of neuron
diagrams and not neuron graphs, we first extend the syntax of the DSEL with a
new operation for computing the causal semantics of \emph{all} diagrams that
can be generated from a graph, and returning these as a list.

\begin{code}
allCauses :: NV a => G a -> [Causes a]
allCauses = map causes . allDiagrams
\end{code}

\noindent
Now we can use this new operation to confirm that every corresonding diagram
generated from each graph has the same causal semantics.

\begin{code}
> allCauses trumpG == allCauses (graph byRank)
True
\end{code}

\noindent
To reiterate the point from Section~\ref{sec:model}, the equivalence of the
causal semantics \emph{does not} follow from the equivalence of the firing
semantics (the graphs of \cod{majorOrders} and \cod{trump} serve as
counterexamples).  Rather, this was simply a feature designed into the
particular representations presented here.
Often it is difficult to determine the causal semantics of a diagram just by
looking at it, especially in the presence of non-standard neurons like the
\cod{ByRank} neuron used above.
With the direct language support provided by our DSEL for viewing and comparing
a neuron diagram's causes, we were able to quickly and reliably confirm the
equivalence of the two representations above, something that would have been
time-consuming and error-prone otherwise.

Support for non-boolean values is a straightforward but significant extension
to the language of neuron diagrams.  Our causal reasoning algorithm easily
accommodates this extension, and our DSEL enables causation researchers to
either adapt existing neuron types to non-boolean domains or create new types
of neurons that operate on these new values.
Other languages for causal reasoning also support non-boolean values
\cite{HP05}, but these scenarios have been simulated in neuron diagrams only
awkwardly, for example, by using several boolean neurons to represent a single
non-boolean event \cite{Hitchcock07}.
One of the main strengths of neuron diagrams is their explanatory power
compared to other representations, but this strength is diminished by the
modeling contortions imposed by the constraints of boolean-only values.  With
the extensions described in this paper and supported by this DSEL, we can lift
these constraints, promoting the creation of direct and readable neuron
diagrams.

\mysection{Neuron Diagrams as Explanations}
\label{sec:expl}

Throughout this paper we have presented only very small neuron diagrams.  These
are toy examples in the sense that they have been chosen to demonstrate
particular aspects of the DSEL.  However, they are also highly representative
of the size and nature of neuron diagrams actually used in the philosophy
literature.  Neuron diagrams are not used for identifying causes in complex
networks of events, but rather for presenting and explaining simple structures
that illustrate some tricky aspect of causation or make a particular point. In
other words, neuron diagrams are a language for toy examples!

In this section we will introduce two (slightly) more substantial examples to
demonstrate the utility of neuron diagrams as explanations and to discuss the
role of neuron diagrams in causation research.

\mysubsection{Explanations of Logic Puzzles}
\label{sec:logic}

We begin by illustrating the explanatory value of representing causal
relationships directly, as edges between nodes. As a motivating example,
consider the following boolean logic puzzle.%
\footnote{From
\url{http://www.cs.princeton.edu/courses/archive/spr06/cos116/COS_116_HW_3.pdf}}
\begin{quote}
{\sl Matt will go to the party if John and Brian go. Brian will go if Karen goes or
Sue doesn't~go. Sue will go if John doesn't. Karen will go if Sue does. When
does Matt go to the party?}
\end{quote}
All causal relationships in the story are encoded concisely and unambiguously
in the above description, but the solution is not obvious and so this
representation's explanatory value is low.

\begin{figure}[t]
\hfill
\begin{figcode}
if\USCORE\   n = StimBy  [n]
ifNot n = UnstimBy [n]
\end{figcode}
\hfill\vline\hfill
\begin{figcode}
ifAny ns = StimBy ns
ifAll ns = Thick (length ns) ns
\end{figcode}
\hfill{}
\caption{Neuron descriptions for conditional and quantified conditional logic operators.}
\label{fig:logic}
\end{figure}

In order to represent this problem as a neuron diagram, we first define a
mini-DSL for encoding encoding conditional logic statements in neuron diagrams,
given in Figure~\ref{fig:logic}. The \cod{if\USCORE} and \cod{ifAny} constructs
produce basic neuron descriptions with stimulating edges. The \cod{ifNot}
construct produces a neuron that will fire only if its predecessor does
\emph{not} fire, represented by an unstimulating edge (hollow arrowhead).
Finally, the \cod{ifAll} construct produces a neuron that fires only if all of
its predecessors also fire, using a thick neuron. Thick neurons and
unstimulating edges were described in Section~\ref{sec:ndinst}.

Using this mini-DSL, plus the {\ttsize\verb#:||:#} operator from
Section~\ref{sec:dec}, we can almost directly translate the above description
of the puzzle into the following neuron graph.

\begin{code}
party = Graph [matt]
  where
    matt  = "Matt"  :# ifAll [john,brian]      
    brian = "Brian" :# if_ karen :||: ifNot sue
    sue   = "Sue"   :# ifNot john              
    karen = "Karen" :# if_ sue                 
    john  = "John"  :# Input
\end{code}

\noindent
The neuron representing John is encoded as an input neuron since it has no
predecessors.

\begin{figure}[t]
\centering
  {\includegraphics[scale=\figscale]{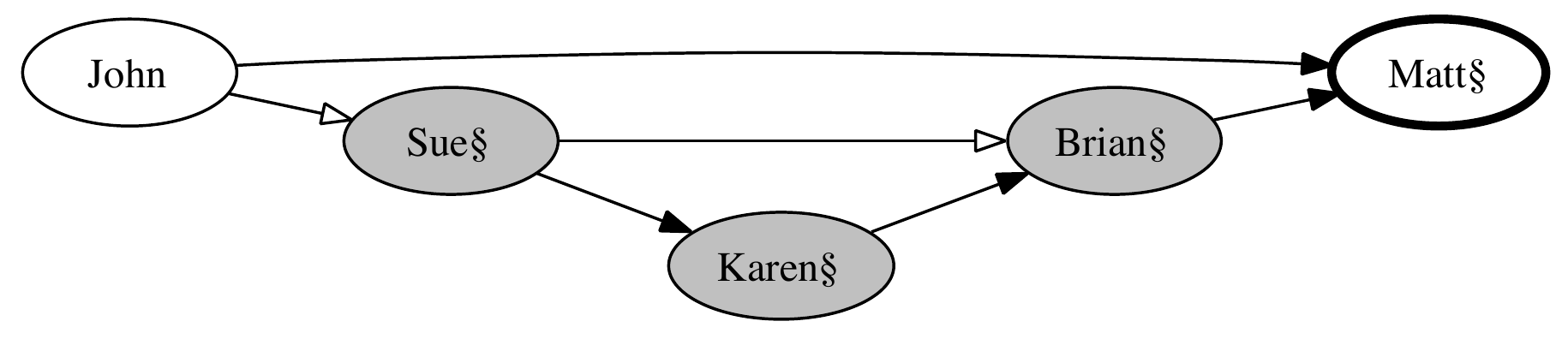}}
  {\includegraphics[scale=\figscale]{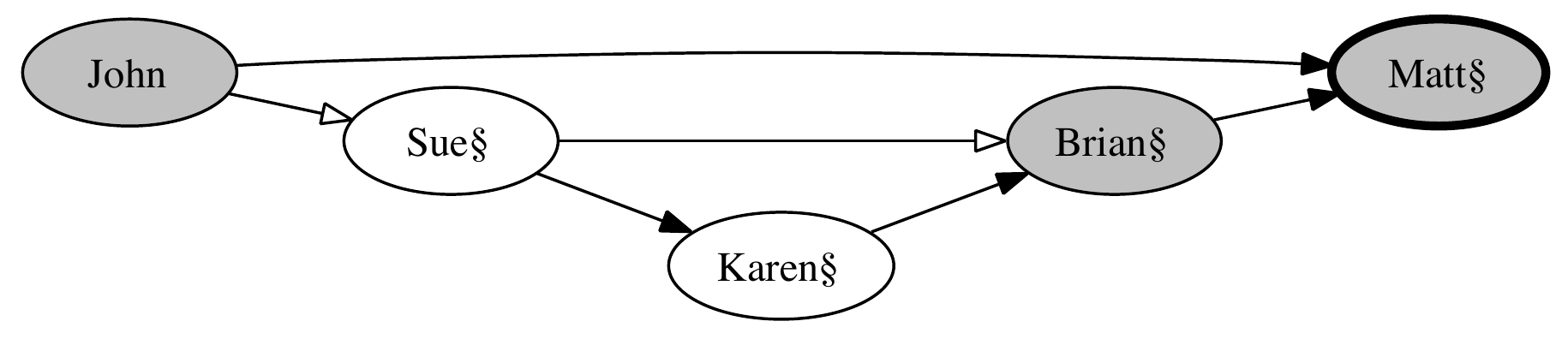}}
\caption[]{Diagrams for the two scenarios encoded in the party logic puzzle.}
\label{fig:party}
\end{figure}

Using this graph we can generate the two diagrams shown in
Figure~\ref{fig:party}. From these diagrams it is immediately clear that
whether or not Matt goes to the party depends counterfactually only on whether
John goes to the party, and we can confirm this by examining the causes of
the diagrams.

\begin{code}
> allCauses party
[John:False ==> Matt:False,John:True ==> Matt:True]
\end{code}

\noindent
Additionally, by showing the causal relationships directly, determining which
events affect and are affected by other events is a \emph{local} operation. For
example, by looking only at the neuron representing Brian and its neighbors, we
can see that Brian is influenced in his decision to attend by Sue and Karen,
and that his decision influences Matt's. In either textual representation, the
indirection created by naming forces us to scan the \emph{entire} model in order to
get the same information, making the operation more difficult.

Finally, by looking at the diagrams we can see at a glance who will attend the
party in either scenario. This requires much more effort when looking at
either the original description or the DSEL representation, since these
representations abstract away from particular instances of the story.

In this example, the mapping from story to neuron diagram was unambiguous.  In
real scenarios things are messier. The value of neuron diagrams then is not in
identifying the correct causes in a story, but rather in comparing different
ways of modeling the story and the different causes they produce.

\mysubsection{The Assassination of James A.\ Garfield}
\label{sec:garfield}

In 1881, U.S.\ President James A.\ Garfield was shot in the back by a rejected
office-seeker named Charles J.\ Guiteau.%
\footnote{\url{http://en.wikipedia.org/wiki/Assassination_of_James_A._Garfield}}
The bullet lodged in Garfield's spine, critically wounding him.  Doctors
believed his recovery depended on removing the bullet but were unsuccessful in
several attempts. Garfield never recovered, dying 11 weeks later from
infections contracted from doctors probing for the bullet with unwashed hands.
During his trial, Guiteau famously argued, ``the doctors killed Garfield, I
just shot him''.

The question, of course, is who caused Garfield's death?
The answer is that it depends on how we model the story! We can construct
convincing neuron diagrams where the doctors are the only cause, where Guiteau
is the only cause, and several variants where both are causes.
This demonstrates the fundamental unsuitability of neuron diagrams for
objectively identifying causes in real-world situations---modeling a story
essentially amounts to encoding a preconceived set of causes into a diagram.
Rather than being a weakness, however, this is exactly the point.
Neuron diagrams are a tool for representing these (preconceived) causal
relationships succinctly so that they can be shared, explained, and justified.
Our DSEL supports this process by providing a programmatic way to generate
diagrams and to confirm that the encoded causes are actually those that the
creator intended.

\begin{figure}[t]
\centering
\subcaptionbox{\label{fig:garfield:savable}}
  {\includegraphics[scale=0.525]{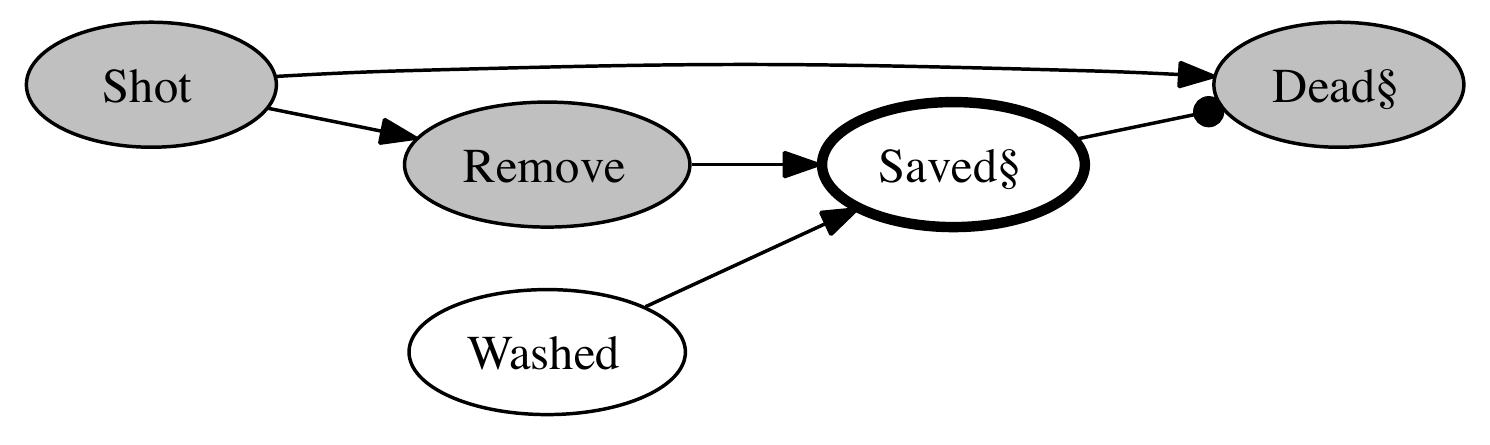}}
\subcaptionbox{\label{fig:garfield:fatal}}
  {\includegraphics[scale=0.525]{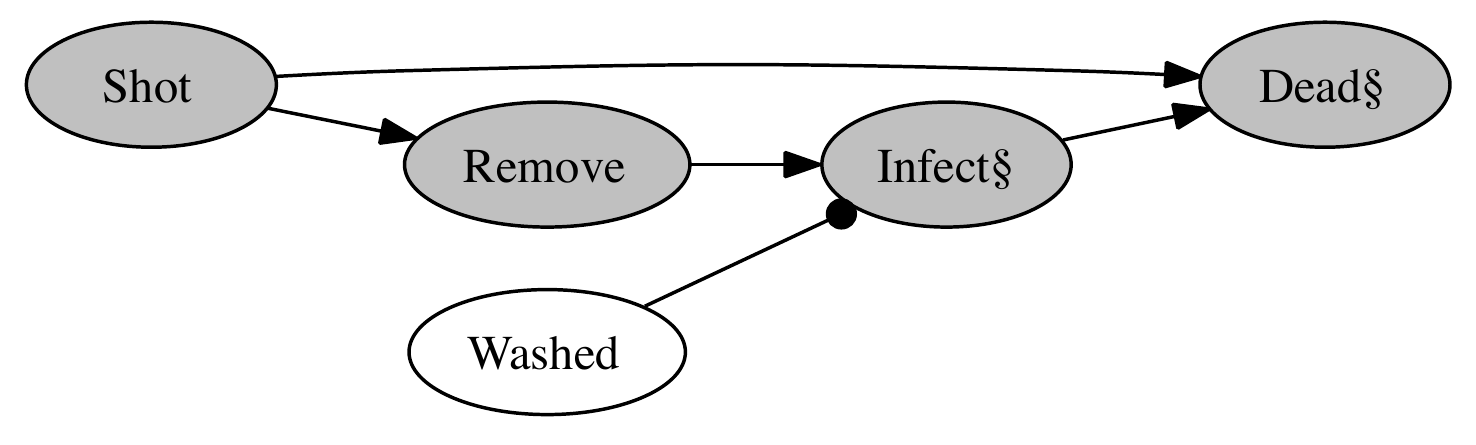}}
\caption[]{Neuron diagrams describing the death of James A.\ Garfield. In
\sref*{fig:garfield}{savable} the doctors' attempts to remove the bullet could
have saved Garfield's life. In \sref*{fig:garfield}{fatal} their attempts were
futile.}
\label{fig:garfield}
\end{figure}

Here we will consider a subtle distinction between two cases in which both
Guiteau and the doctors are identified as causes of Garfield's death. The
neuron diagrams for these cases are shown in Figure~\ref{fig:garfield}. In both
diagrams, the \Shot\ neuron represents Guiteau shooting the president, \Remove\
represents the doctors attempting to remove the bullet, \Washed\ represents
whether the doctors washed their hands, and \Dead\ represents the death of
Garfield. 
\Shot\ and \Washed\ are typical input neurons which we assume are bound to the
Haskell identifiers \cod{shot} and \cod{washed}. The doctors will only attempt
to remove the bullet if Garfield has been shot, so we represent the \Remove\
neuron in our DSEL as follows.

\begin{code}
remove = "Remove" :# if_ shot `IsKind` Action
\end{code}

\noindent
We mark this neuron as an action in order to prevent the benefits of this
action from being credited to the shot itself, similar to the hiker-boulder
example from Section~\ref{sec:cause}.

In Figure~\sref{fig:garfield}{savable}, the \Saved\ neuron
represents the possibility that the doctors could have saved Garfield. In order
to be saved, the doctors must attempt to extract the bullet with clean
hands; if saved, Garfield's death will be prevented. We define this
diagram in the DSEL as follows, where \cod{unless} decorates a neuron
description with a single inhibitory edge (\cod{unless~d~n~=~InhibBy~d~[n]}).


\begin{code}
savable = diagram [dead] `WithInputs` [True,False]
  where
    dead  = "Dead"  :# if_ shot `unless` saved
    saved = "Saved" :# ifAll [remove,washed]
\end{code}

\noindent
If the doctors do not wash their hands, Garfield will not be saved since he
will die from infection.

In Figure~\sref{fig:garfield}{fatal} we represent the risk of infection more
explicitly, by a neuron that will fire if the doctors attempt to remove the
bullet but do not wash their hands.
This model also suggests that the wound itself was fatal, that
removing the bullet would not have saved Garfield (as some historians believe).

\begin{code}
fatal = diagram [dead] `WithInputs` [True,False]
  where
    dead   = "Dead"   :# ifAny [shot,infect]
    infect = "Infect" :# if_ remove `unless` washed
\end{code}

\noindent
These diagrams represent different but equally valid interpretations of the
death of Garfield, but as we can see through cause inference, the causes they
encode are very different.

\begin{code}
> causes savable
Shot:True&Washed:False ==> Dead:True
> causes fatal
Shot:True | Removed:True&Washed:False ==> Dead:True
\end{code}

\noindent
If we believe that Garfield was savable, then his death was caused by the
combination of Guiteau's shot and the doctors' unwashed hands. However, if
we think that Garfield's wound was already fatal, then Garfield's death is
overdetermined. In this case, the shot alone is a sufficient cause of
Garfield's death, as is the attempt to remove the bullet with unwashed hands.

These diagrams represent just two of many ways that Garfield's death could be
modeled.  The causes produced by any single account are merely a reflection of
the assumptions that went into the construction of that model, but by exploring
the space of causal models we can reflect on those assumptions and their impact
on the inferred causes, leading to a deeper understanding of the situation and
the nature of causation itself.
The DSEL presented in this paper is a tool for rapidly generating, visualizing,
and analyzing causal models, and therefore supports this process directly.

\mysection{Related Work}
\label{sec:rw}

While the focus of our DSEL and of neuron diagrams are on reasoning about
deterministic causation, reasoning about causation under uncertainty has been
an important topic in AI research. For an overview of the vast literature in
this area, see the recently updated version of Judea Pearl's monograph
\cite{Pearl09}.

Since the introduction of the visual language of neuron diagrams by David Lewis
\cite{Lewis87}, neuron diagrams have been used extensively to investigate and
explain causation in particular situations and to study the nature of causation
itself.  One of the attractive features of neuron diagrams is that they provide
an immediate visual representation for counterfactual reasoning.  The
relationship of counterfactuals to causation was first expressed by 18th
century philosopher David Hume \cite{Hume1748}, but the first fully developed
theory of causation in terms of counterfactuals was introduced by Lewis
\cite{Lewis73}.  In the same work, Lewis also develops the notion of causal
chains, central to our causal reasoning algorithm.


Most research on counterfactually determined causation has focused on ``token''
causation \cite{Hitchcock01}, which in neuron diagrams corresponds to causes
involving only a single neuron.  An exception to this is the work of computer
scientists Halpern and Pearl \cite{HP05}, who developed the structural
equations model, a more expressive but mostly non-visual language for
describing and reasoning about causality.  Their idea of ``general'' causation,
where sets of events can act as causes, has since propagated back into
philosophical analyses by, for example, Hitchcock \cite{Hitchcock01}, Hall
\cite{Hall07}, and Woodward \cite{Woodward03}.
Interestingly, none of these approaches distinguishes between conjunctions and
disjunctions of events in causes, which provides a more accurate description of
causes in certain situations, as we have shown in \cite{EW10hcc}.

Although structural equations have their own impoverished graphical notation
called ``causal graphs'', it is much less expressive than neuron diagrams
\cite{Pearl09}.
%
%
Causal graphs are equivalent to a neuron graph in which all neurons and edges
are the same shape and style.
This notation is also peripheral to the language of structural
equations---causal graphs document the relationships encoded in a structural
equations model, but it is entirely possible to use structural equations without ever
considering causal graphs.

Despite their wide-spread use and popularity, neuron diagrams have come under
criticism.  In particular, Hitchcock \cite{Hitchcock07} details several
perceived limitations of the language.  The most critical of these are that the
language's inherent extensibility makes neuron diagrams difficult to
reason about and not enumerable.  Enumerability is important, he argues,
to compare a neuron diagram to possible alternatives.
Our formalization of neuron diagrams and their semantics resolves both of these
concerns \cite{EW10hcc}.  Furthermore, our extension of the language to
distinguish between action and law neurons has allowed us to develop a cause
inference algorithm that produces more accurate results than any previous
approach.
The work presented here provides an implementation of these theoretical
developments, supporting practical work with neuron diagrams.  It 
allows philosophers and other causation researchers to easily generate
diagrams, systematically enumerate all cases for a class of causal situations
given by a neuron graph, and most importantly, to automatically and accurately
infer the causes encoded in a diagram.
%

In other previous work, we have developed a DSEL 
for creating explanations of probabilistic reasoning problems \cite{EW09dsl}.
This language relies fundamentally on the principle of causation and provides
constructs to combine events into stories.  Unlike neuron diagrams, however,
these stories have a mostly linear form.
Moreover, no operations exist in the language to infer the causes of
events.

\mysection{Conclusions and Future Work}
\label{sec:fw}

In this paper we have presented a DSEL to support causation research.  The
specific contributions of this work include: (1) an implementation of our
formal model of neuron diagrams and our cause inference algorithm, making this
previously theoretical work practically accessible and applicable for causation
researchers; (2) an extension of the neuron diagram language to operate on
non-boolean values; (3) a domain-specific language that supports the easy
creation of neuron diagrams that is easily extended by new types of neurons and
values; and (4) a supporting library of operations for manipulating, querying,
and altering neuron graphs and diagrams.


In addition to the applications described in this paper, we can imagine several
more advanced applications of our language.  For example, the language can be
used as a platform for testing and comparing alternative cause inference
algorithms, or as a platform for research on explanations by comparing which
equivalent neuron diagrams are easier for people to understand.  The extension
of neuron diagrams to arbitrary bounded and enumerable data types could also
lead to many unexpected applications.


For our own future work, we intend to extend this DSEL with a query language
for neuron diagrams.  Such a language would provide, at least, generators for
neurons, neuron graphs, and neuron diagrams, and filters for identifying graphs
with certain effects or diagrams that encode particular causes.
Such a language would be a boon to causation research, allowing one to easily
compare all causal structures that share important properties, and to rigorously
test an idea against a battery of relevant test cases.

\bibliographystyle{eptcs}
\bibliography{me,explain,sprsh,se,fp}

\end{document}

%% file: ndd.tex
\newcommand{\OB}[1]{\ensuremath{#1}}

\newcommand{\viGen}[4]{\OB{#1_{#2}#4\ldots#4#1_{#3}}}
\newcommand{\vij}[3]{\viGen{#1}{#2}{#3}{,}}
\newcommand{\vi}[2]{\vij{#1}{1}{#2}}
\newcommand{\vn}[2][-]{\vi{#2}{\if#1-n\else#1\fi}}

\newcommand{\name}[1]{\textit{#1}}

\newcommand{\Dead}{\name{Dead}}

\newcommand{\Boulder}{\name{Boulder}}
\newcommand{\Duck}{\name{Duck}}

\newcommand{\lawSym}{\textrm{\textsection}}

